\documentclass[a4paper,12pt]{article}
\usepackage{amsmath}
\usepackage{amssymb}
\usepackage{amsthm}
\usepackage[dvips]{graphicx}

\begin{document}

\baselineskip=22pt

\author{V.D. Lakhno}
\title{Superconducting properties of 3D low-density TI-bipolaron gas.}
\date
{Institute of Mathematical Problems of Biology RAS, \\
 \it{142290, Pushchino, Moscow region, Russia.}}
\maketitle

\baselineskip=22pt

\begin{abstract}
Consideration is given to thermodynamical properties of a three-dimensional Bose-condensate of translation-invariant bipolarons (TI-bipolarons).  The critical temperature of transition, energy, heat capacity and the transition heat of TI-bipolaron gas are calculated. The results obtained are used to explain experiments on high-temperature superconductors.
\end{abstract}

Victor Lakhno. \verb"email: lak@impb.psn.ru"

\section{Introduction.}

The theory of superconductivity is one of the finest and old subject-maters of condensed matter physics which involves both macroscopic and microscopic theories as well as derivation of macroscopic equations of the theory from microscopic description ~\cite{1}.
In this sense the theory was thought to be basically completed and its further development was to have been concerned with further detalization and consideration of various special cases.

The situation changed when the high-temperature superconductivity (HTSC) was discovered ~\cite{2}. Surprisingly, it was found that in oxide ceramics, the correlation length is some orders of magnitude less than that in traditional metal superconductors while the ratio of the energy gap to the temperature of superconducting transition is much greater ~\cite{3}. The current status of research can be found in books and reviews ~\cite{4}-\cite{15}.

Today the main problem in this field is to develop a microscopic theory capable of explaining experimental facts which cannot be accounted for by the standard BCS theory.  One might expect that development of such a theory would not affect the macroscopic theory based on phenomenological approach.

With all the variety of modern versions of HTSC microscopic descriptions: phonon, plasmon, spin, exciton and other mechanisms, the central point of constructing the microscopic theory is the effect of electron pairing (Cooper effect). In what follows such a “bosonization” of electrons provides the basis for the description of their superconducting condensate.

The phenomenon of pairing, in a broad sense, is considered as arising of bielectron states, while in a narrow sense, if the description is based on phonon mechanism, it is treated as formation of bipolaron states. For a long time this view was hindered by a large correlation length or the size of Cooper pairs in BCS theory. For the same reason, over a long period, the superconductivity was not viewed as a boson condensate (see footnote at p. 1177 in ~\cite{16}). A significant reason of this lack of understanding was a standard idea that bipolarons are very compact particles.

The most dramatic illustration is the use of the small-radius bipolaron (SRB) theory to describe HTSC ~\cite{10}, \cite{17}, \cite{16a}. It implies that a stable bound bipolaron state is formed at one node of the lattice and subsequently such small-radius bipolarons are considered as a gas of charged bosons (as a variant individual SRP are formed and then are considered within BCS of creation of the bosonic states). Despite the elegance of such a picture, its actual realization for HTSC comes up against inextricable difficulties caused by impossibility to meet antagonistic requirements. On the one hand, the constant of electron-phonon interaction (EPI) should be large for bipolaron states of small radius to form. On the other hand, it should be small for the bipolaron mass (on which the superconducting temperature depends ~\cite{17a}-\cite{22}) to be small too. Obviously, the HTSC theory based on SRP concept which uses any other (non-phonon) interaction mechanisms mentioned above will run into the same problems.

Alternatively, in describing HTSC one can believe that the role of a fundamental charged boson particle can be played by large-radius bipolarons (LRB) ~\cite{23}-\cite{27}. Historically just this assumption was made by Ogg and Schafroth ~\cite{23}, ~\cite{28} long before the development of the SRP theory. When viewing Cooper pairs as a peculiar kind of large-radius bipolaron states, one might expect that the LRP theory should be used to solve the HTSC problem.

As pointed out above, the main obstacle to consistent use of the LRP theory for explaining high-temperature superconductivity was an idea that electron pairs are localized in a small region, the constant of electron-phonon coupling should be large and, as a consequence, the effective mass of electron pairs should be large.

In the light of the latest advances in the theory of LRP and LRB, namely, in view of development of an all-new concept of delocalized polaron and bipolaron states – translation-invariant polarons (TI-polarons) and bipolarons (TI-bipolarons)  ~\cite{29}-\cite{35}, it seems appropriate to consider their role in the HTSC theory in a new angle.

We recall the main results of the theory of TI-polarons and bipolarons obtained in ~\cite{29}-\cite{35}. Notice that consideration of just electron-phonon interaction is not essential for the theory and can be generalized to any type of interaction.

In what follows we will deal only with the main points of the theory important for the HTSC theory. The main result of papers ~\cite{29}-\cite{35}  is construction of delocalized polaron and bipolaron states in the limit of strong electron-phonon interaction. The theory of TI-bipolarons is based on the theory of TI-polarons ~\cite{29}, ~\cite{30} and retains the validity of basic statements proved for TI-polarons. The chief of them is the theorem of analytic properties of the ground state of a TI-polaron (accordingly TI-bipolaron) depending on the constant of electron-phonon interaction $\alpha$. The main implication of this statement is the absence of a critical value of the EPI constant $\alpha_c$, below which the bipolaron state becomes impossible since it decays into independent polaron states. In other words, if there exists a value of $\alpha_c$, at which the TI-state becomes energetically disadvantageous with respect to its decay into individual polarons, then nothing occurs at this point but for $\alpha<\alpha_c$ the state becomes metastable.
Hence, over the whole range of $\alpha$ variation we can consider TI-polarons as charged bosons capable of forming a superconducting condensate.

Another important property of TI-bipolarons is the possibility of changing the correlation length over the whole range of $[0,\infty]$ depending on the Hamiltonian parameters ~\cite{32}. Hence, it can be both much larger (as is the case in metals) and much less than the characteristic size between the electrons in an electron gas (as happens with ceramics).

A detailed description of the theory of TI-polarons and bipolarons and description of their various properties is given in review ~\cite{35}.

An outstandingly important property of TI-polarons and bipolarons is the availability of an energy gap between their ground and excited states (\S3).

The above-indicated characteristics can be used to develop a microscopic HTSC theory on the basis of TI-bipolarons.

The paper is arranged as follows. In \S2 we take Pekar-Froehlich Hamiltonian for a bipolaron as an initial Hamiltonian. The results of three canonical transformations, such as Heisenberg transformation, Lee-Low-Pines transformation and that of Bogolyubov-Tyablikov are briefly outlined. Equations determining the TI-bipolaron spectrum are derived.

In \S3 we analyze solutions of the equations for the TI-bipolaron spectrum. It is shown that the spectrum has a gap separating the ground state of a TI-bipolaron from its excited states which form a quasicontinuous spectrum. The concept of an ideal gas of TI-bipolarons is substantiated.

With the use of the spectrum obtained, in \S4 we consider thermodynamic characteristics of an ideal gas of TI-bipolarons. For various values of the parameters, namely phonon frequencies, we calculate the values of critical temperatures of Bose condensation, latent heat of transition into the condensed state, heat capacity and heat capacity jumps at the point of transition.

In \S5 we discuss the nature of current states in Bose-condensate of TI-bipolarons. It is shown that the transition from a currentless state to a current one is sharp.

In \S6 the results obtained are compared with the experiment.

In \S7 we consider the problems of expanding the theory which would enable one to make a more detailed comparison with experimental data on HTSC materials.

In \S8 we sum up the results obtained.

\section{Pekar-Froehlich Hamiltonian. Canonical transformations.}

Following ~\cite{31}-\cite{35}, in describing bipolarons we will proceed from Pekar-Froehlich Hamiltonian:
\begin{eqnarray}\label{1}
    H=-\frac{\hbar^2}{2m^*}\Delta_{r_1}-\frac{\hbar^2}{2m^*}\Delta_{r_2}+
		\sum_k{\hbar\omega^0_k a^+_k a_k}+U\left(\left|\vec{r}_1-\vec{r}_2\right)\right|+\\ \nonumber
		\sum_k\left(V_k e^{i\vec{k}\vec{r}_1}a_k+V_k e^{i\vec{k}\vec{r}_2}a_k+H.c.\right),\ \ \
		U\left(\left|\vec{r}_1-\vec{r}_2\right|\right)=
		\frac{e^2}{\epsilon_{\infty}\left|\vec{r}_1-\vec{r}_2\right|}
\end{eqnarray}
where $\vec{r}_1$ $\vec{r}_2$ are coordinates of the first and second electrons, respectively;
$a^+_k$, $a_k$  are operators of the birth and annihilation of the field quanta with energy $\hbar\omega^0_k=\hbar\omega_0$; $m^*$ is the electron effective mass;
the quantity $U$ describes Coulomb repulsion between the electrons; $V_k$ is the function of the wave vector $k$:
\begin{multline}\label{2}
     V_k=\frac{e}{k}\sqrt{\frac{2\pi\hbar\omega_0}{\tilde{\epsilon}V}}=
		\frac{\hbar\omega_0}{ku^{1/2}}\left(\frac{4\pi\alpha}{V}\right)^{1/2},
		u=\left(\frac{2m^*\omega_0}{\hbar}\right)^{1/2},\ \
		\alpha=\frac{1}{2}\frac{e^2u}{\hbar\omega_0\tilde{\epsilon}},\\
		\tilde{\epsilon}^{-1}=\epsilon^{-1}_{\infty}-\epsilon^{-1}_{0},
\end{multline}
where $e$ is the electron charge; $\epsilon_{\infty}$ and $\epsilon_0$ are high-frequency and static dielectric permittivities; $\alpha$ is the constant of electron-phonon interaction; $V$ is the system’s volume.

In the system of the center of mass Hamiltonian \eqref{1} takes the form:
\begin{multline}\label{3}
     H=-\frac{\hbar^2}{2M_e} \Delta_R-\frac{\hbar^2}{2\mu_e} \Delta_r+\sum_k{\hbar\omega^0_k a^+_k a_k}+U(|\vec{r}|)+
		\sum_k{2V_k \text{cos} \frac{\vec{k}\vec{r}}{2}}\left(a_k e^{i\vec{k}\vec{R}}+H.c.\right),\\
		\vec{R}=(\vec{r}_1+\vec{r}_2)/2,\ \ \vec{r}=\vec{r}_1-\vec{r}_2,\ \ M_e=2m^*,\ \ \mu_e=m^*/2
\end{multline}
In what follows in this section we will believe  $\hbar=1$, $\omega^0_k=1$, $M_e=1$ (accordingly $\mu_e=1/4$).

The coordinates of the center of mass $\vec{R}$ can be excluded from Hamiltonian \eqref{3} using Heisenberg’s canonical transformation ~\cite{36}:
\begin{equation}\label{4}
    S_1=\text{exp}\left\{-i\sum_k\vec{k}a^+_k a_k\right\}\vec{R},\ \ \
		S^{-1}_1a_kS_1=a_ke^{-i\vec{k}\vec{R}},\ \ \
		S^{-1}_1a^+_kS_1=a^+_ke^{i\vec{k}\vec{R}}.
\end{equation}
Accordingly, the transformed Hamiltonian will be written as:
\begin{equation}\label{5}
    \tilde{H}=S^{-1}_1HS_1=-2\Delta_r+U(|\vec{r}|)+\sum_ka^+_ka_k+
		\sum_k2V_k\text{cos}\frac{\vec{k}\vec{r}}{2}(a_k+a^+_k)+
		\frac{1}{2}\left(\sum_k\vec{k}a^+_ka_k\right)^2
\end{equation}
From \eqref{5} it follows that the exact solution of the bipolaron function is determined by the wave function $\psi(r)$, which contains only relative coordinates $r$ and, therefore, is translation-invariant.

Averaging of $\tilde{H}$ over $\psi(r)$ yields the Hamiltonian $\bar{H}$:
\begin{multline}\label{6}
     \bar{H}=\frac{1}{2}\left(\sum_k\vec{k}a^+_ka_k\right)^2+\sum_k a^+_k a_k +
		\sum_k\bar{V}_k(a_k+a^+_k)+\bar{T}+\bar{U},\\
		\bar{V}_k=2V_k\left\langle \Psi\left|\text{cos}\frac{\vec{k}\vec{r}}{2}\right|\psi\right\rangle,\ \
		\bar{U}=\left\langle \Psi\left|U(r)\right|\Psi\right\rangle,\ \
		\bar{T}=-2\left\langle \Psi\left|\Delta_r\right|\Psi\right\rangle.
\end{multline}
Eq. \eqref{6} suggests that the bipolaron Hamiltonian differs from the polaron one in that in the latter the quantity  $V_k$ is replaced by $\bar{V}_k$ and the constants  $\bar{T}$, $\bar{U}$ are added.

With the use of Lee-Low-Pines canonical transformation ~\cite{37}:
\begin{equation}\label{7}
    S_2=\text{exp}\left\{\sum_kf(k)(a^+_k-a_k)\right\},
\end{equation}
where $f_k$ are variational parameters having the sense of the distance by which the field oscillators are displaced from their equilibrium positions:
\begin{equation}\label{8}
    S^{-1}_2a_kS_k=a_k+f_k,\ \ \ S^{-1}_2a^+_kS_k=a^+_k+f_k,
\end{equation}
for Hamiltonian $\tilde{\tilde{H}}$:
\begin{equation}\label{9}
    \tilde{\tilde{H}}=S^{-1}_2\bar{H}S_2,
\end{equation}
we get:
$$ \tilde{\tilde{H}}=H_0+H_1,$$
\begin{multline}\label{10}
    H_0=2\sum_k\bar{V}_kf_k+\sum_kf_k^2+\frac{1}{2}\left(\sum_k\vec{k}f^2_k\right)^2+
		\mathcal{H}_0+\bar{T}+\bar{U},\\
		 \mathcal{H}_0=\sum_k\omega_ka^+_ka_k+\frac{1}{2}\sum_{k,k'}\vec{k}\vec{k'}f_kf_{k'}
		(a_ka_{k'}+a^+_ka^+_{k'}+a^+_ka_{k'}+a^+_{k'}a_k),\\
		\omega_k=1+\frac{k^2}{2}+\vec{k}\sum_{k'}\vec{k}'f^2_{k'}.
\end{multline}
Hamiltonian  $H_1$ contains terms linear, threefold and fourfold in the birth and annihilation operators. Its explicit form is given in  ~\cite{29}-\cite{31}.

Then, as is shown in ~\cite{29}-\cite{30}, the use of Bogolyubov-Tyablikov canonical transformation ~\cite{38} for passing on from operators $a^+_k$, $a_k$ to new operators $\alpha^+_k$,$\alpha_k$:
$$a_k=\sum_{k'}M_{1kk'}\alpha_{k'}+\sum_{k'}M^*_{2kk'}\alpha^+_{k'}$$
\begin{equation}\label{11}
     a^+_k=\sum_{k'}M^*_{1kk'}\alpha^+_{k'}+\sum_{k'}M_{2kk'}\alpha_{k'}
\end{equation}
(in which $\mathcal{H}_0$ is a diagonal operator), makes mathematical expectation of $H_1$ equal to zero.

In the new operators $\alpha^+_k$, $\alpha_k$ Hamiltonian \eqref{10} takes on the form  $\tilde{\tilde{\tilde{H}}}$:
$$\tilde{\tilde{\tilde{H}}}=E_{bp}+\sum_k\nu_k\alpha^+_k\alpha_k,$$
\begin{equation}\label{12}
    E_{bp}=\Delta E_r+2\sum_k\bar{V}_kf_k+\sum_kf^2_k+\bar{T}+\bar{U},
\end{equation}
where $\Delta E_r$ is the so-called “recoil energy”.
The general expression for $\Delta E_r=\Delta E_r\left\{f_k\right\}$ was obtained in ~\cite{30}.
Actually, calculation of the ground state energy $E_{bp}$ was performed in ~\cite{34} by minimization of \eqref{12} in $f_k$ and in $\psi$.

Notice that in the theory of a polaron with broken symmetry a diagonalized electron-phonon Hamiltonian has the form of \eqref{12}~\cite{41-Miyake}. This Hamiltonian can be treated as a Hamiltonian of a polaron and a system of its associated renormalized real phonons or as a Hamiltonian whose quasiparticle excitations spectrum is determined by \eqref{12}~\cite{42-Levenson}. In the latter case excited states of a polaron are Fermi quasiparticles.

In the case of a bipolaron the situation is qualitatively different since a bipolaron is a boson quasiparticle whose spectrum is determined by \eqref{12}. Obviously, a gas of such quasiparticles can experience Bose-Einstein condensation (BEC). Treatment of \eqref{12} as a bipolaron and its associated renormalized phonons does not prevent their BEC since maintenance of the number of particles required in this case takes place automatically due to commutation of the total number of renormalized phonons with Hamiltonian \eqref{12}.

Renormalized frequencies $\nu_k$ involved in \eqref{12}, according to ~\cite{29}, ~\cite{30} are determined by the equation for  $s$:
\begin{equation}\label{13}
    1=\frac{2}{3}\sum_k\frac{k^2f^2_k\omega_k}{s-\omega^2_k},
\end{equation}
solutions of which yield the spectrum of $s=\left\{\nu^2_k\right\}$ values.

\section{Energy spectrum of a TI-bipolaron.}

Hamiltonian \eqref{12} is conveniently presented in the form:

\begin{equation}\label{14}
    \tilde{\tilde{\tilde{H}}}=\sum_{n=0,1,2,...}E_{n}\alpha^+_{n}\alpha_{n},
\end{equation}
\begin{equation}\label{15}
    E_{n}=\left\{\begin{array}{rl}
		E_{bp},\ n=0;\\
		\nu_{n}=E_{bp}+\omega_0+\frac{k^2_{n}}{2},\ n\neq0 ;
		\end{array}\right. \
\end{equation}
where in the case of a three-dimensional ionic crystal:
\begin{equation}\label{16}
    k_{n_i}=\pm\frac{2\pi(n_i-1)}{N_{a_i}},\ \ n_i=1,2,...,\frac{N_{a_i}}{2}+1,\ i=x,y,z,
\end{equation}
$N_{a_i}$ - is the number of atoms along the \textit{i}-th crystallographic axis.

Let us prove the validity of the expression for the spectrum \eqref{14}, \eqref{15}. Since operators $\alpha^+_{n}$, $\alpha_{n}$ obey Bose commutation relations:
\begin{equation}\label{17}
    \left[\alpha_{n},\alpha^+_{n'}\right]=\alpha_{n}\alpha^+_{n'}-\alpha^+_{n'}\alpha_{n}=
		\delta_{n,n'},
\end{equation}
they can be considered to be operators of birth and annihilation of TI-bipolarons. The energy spectrum of TI-bipolarons, according to \eqref{13}, is determined by the equation:
\begin{equation}\label{18}
    F(s)=1
\end{equation}
where:
\begin{equation}\label{19}
    F(s)=\frac{2}3{\sum_{n}}\frac{k^2_{n}f^2_{k_n}\omega^2_{k_n}}{s-\omega^2_{k_n}}
\end{equation}
It is convenient to solve equation \eqref{18} graphically (Fig.1):

\begin{center}

\includegraphics{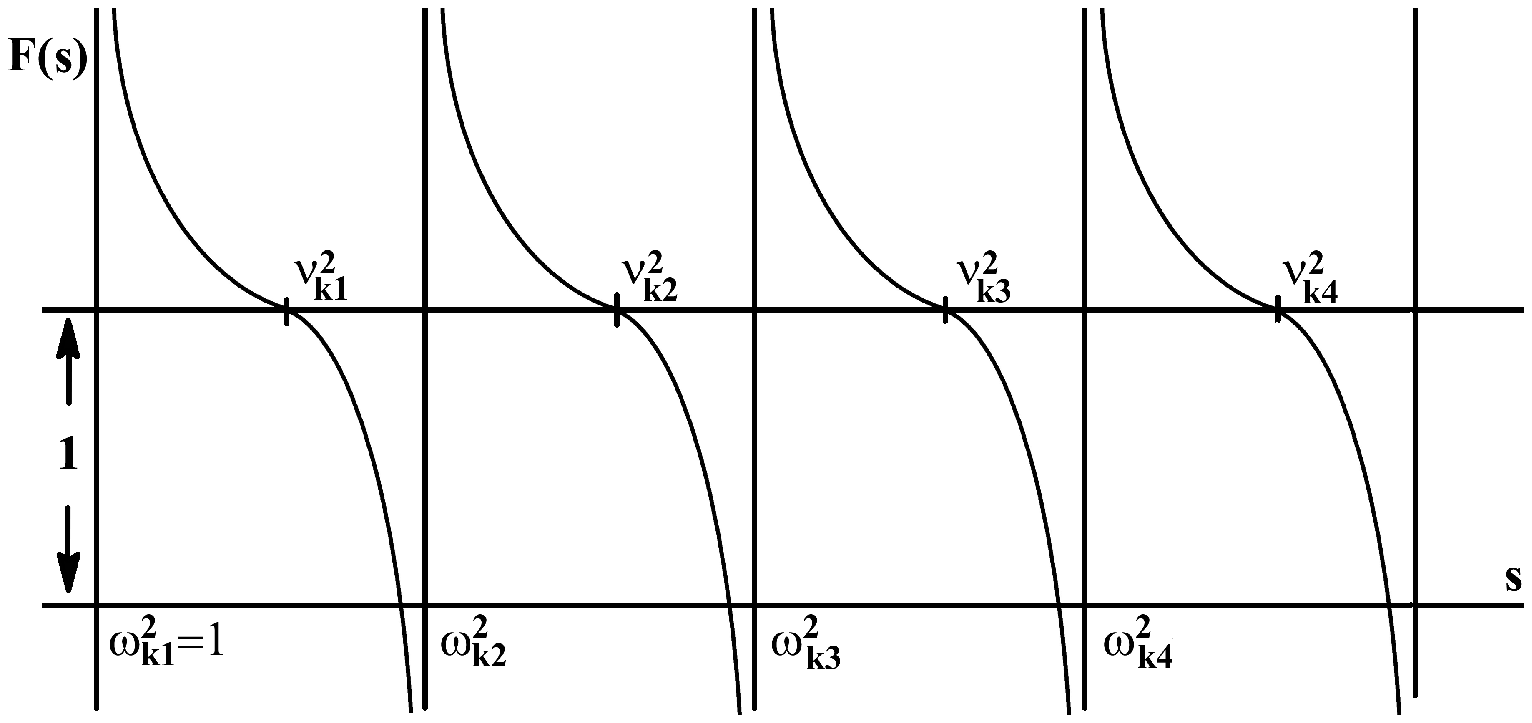}

Fig.1. Graphical solution of equation \eqref{18}.
\end{center}

Fig.1. suggests that the frequencies  $\nu_{k_n}$ (index i is omitted) lie between the frequencies $\omega_{k_n}$ and $\omega_{k_{n+1}}$. Hence, the spectrum  $\nu_{k_n}$ as well as the spectrum $\omega_{k_n}$ are quazicontinuous: $\nu_{k_n}-\omega_{k_n}=O(N^{-1})$, which just proves the validity of \eqref{14}, \eqref{15}.

It follows that the spectrum of a TI-bipolaron has a gap between the ground state $E_{bp}$ and the quasicontinuous spectrum, equal to $\omega_0$.

Below we will consider the case of low concentration of TI-bipolarons in a crystal. Then they can adequately be considered as an ideal Bose gas, whose properties are determined by Hamiltonian \eqref{14}.

\section{Statistical thermodynamics of low-density TI-bipolaron gas.}

Let us consider an ideal Bose-gas of TI-bipolarons which represents a system of $N$ particles occurring in some volume $V$. Let us write $N_0$ for the number of particles in the lower one-particle state and  $N'$ for the number of particles in higher states. Then:
\begin{equation}\label{20}
    N=\sum_{n=0,1,2,...}\bar{m}_{n}=\sum_{n}\frac{1}{e^{(E_{n}-\mu)/T}-1},
\end{equation}

\begin{equation}\label{21}
    N=N_0+N',\ \ N_0=\frac{1}{e^{(E_{bp}-\mu)/T}-1},\ \
		N'=\sum_{n\neq0}\frac{1}{e^{(E_{n}-\mu)/T}-1}.
\end{equation}
In expression $N'$ \eqref{21}, we will perform integration over quasicontinuous spectrum (instead of summation) \eqref{14}, \eqref{15} and assume $\mu=E_{bp}$.  As a result, from \eqref{20}, \eqref{21} we get an equation for determining the critical temperature of Bose-condensation $T_c$:

\begin{equation}\label{22}
    C_{bp}=f_{\tilde{\omega}}\left(\tilde{T}_c\right),
\end{equation}
$$f_{\tilde{\omega}}\left(\tilde{T}_c\right)=\tilde{T}^{3/2}_cF_{3/2}\left(\tilde{\omega}/\tilde{T}_c\right),\ \
F_{3/2}(\alpha)=\frac{2}{\sqrt{\pi}}\int^{\infty}_0\frac{x^{1/2}dx}{e^{x+\alpha}-1},$$
$$C_{bp}=\left(\frac{n^{2/3}2\pi\hbar^2}{M_e\omega^*}\right)^{3/2},\ \
\tilde{\omega}=\frac{\omega_0}{\omega^*},\ \ \tilde{T}_c=\frac{T_c}{\omega^*},$$
where $n=N/V$. The relation between the notation $F_{3/2}$ with other notations is given in the Appendix. Fig.2 shows a graphical solution of equation \eqref{22} for the values of parameters $M_e=2m^*=2m_0$, where $m_0$ is the mass of a free electron in vacuum, $\omega^*=5$ meV ($\approx$58K), $n=10^{21}$ cm$^{-3}$ and the values: $\tilde{\omega}_1=0,2$; $\tilde{\omega}_2=1$; $\tilde{\omega}_3=2$;
$\tilde{\omega}_4=10$, $\tilde{\omega}_5=15$, $\tilde{\omega}_6=20$.

\begin{center}

\includegraphics{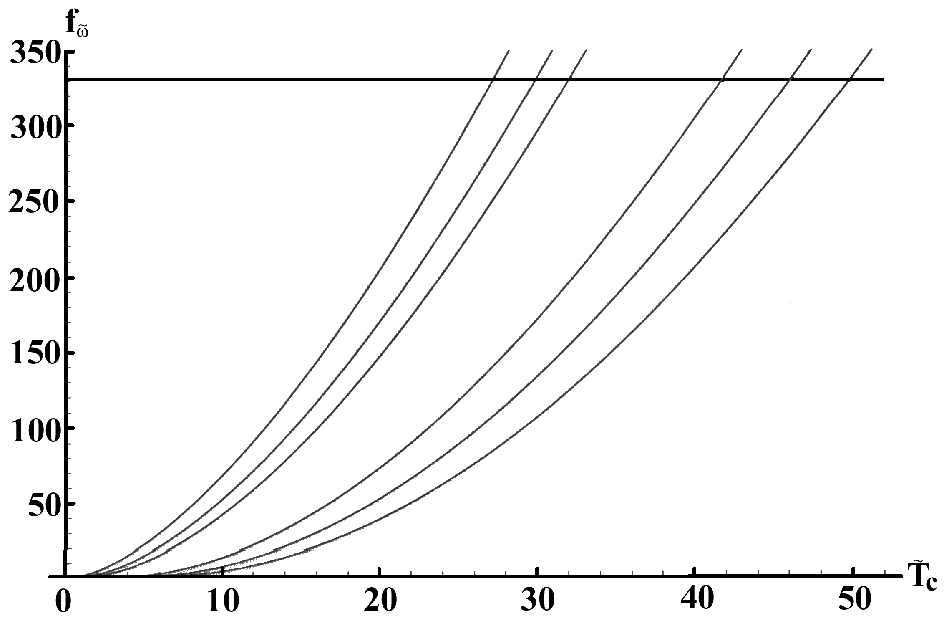}

\end{center}

\begin{center}

\includegraphics{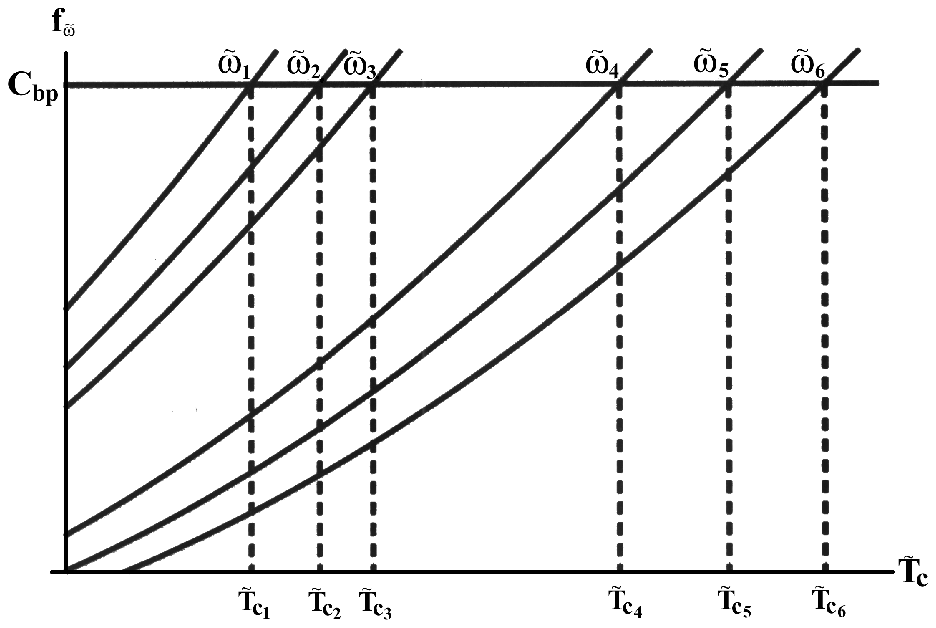}

Fig. 2. Solutions of equation \eqref{22} with $C_{bp}=331,35$ and $\tilde{\omega}_i=\left\{0,2; 1; 2; 10; 15; 20\right\}$,\\
which correspond to
$\tilde{T}_{c_i}$: $\tilde{T}_{c_1}=27,325$; $\tilde{T}_{c_2}=30,0255$; \\
$\tilde{T}_{c_3}=32,1397$;
$\tilde{T}_{c_4}=41,8727$; $\tilde{T}_{c_5}=46,1863$; $\tilde{T}_{c_6}=49,9754$.
\end{center}
It is seen from fig.2 that the critical temperature grows with increasing phonon frequency $\omega_0$.
The relations of critical temperatures $T_{ci}/\omega_{0i}$ corresponding to the chosen parameter values are given in Table I. Table I suggests that the critical temperature of a TI-bipolaron gas is always higher than that of ideal Bose-gas (IBG).
It is also evident from fig.2 that an increase in the concentration of TI-bipolarons $n$ will lead to an increase in the critical temperature, while a gain in the electron mass $m^*$ – to its decrease.  For $\tilde{\omega}=0$ the results go over into the limit of IBG. In particular, \eqref{22} for $\tilde{\omega}=0$, yields the expression for the critical temperature of IBG:
\begin{equation}\label{23}
    T_c=3,31\hbar^2n^{2/3}/M_e
\end{equation}
It should be stressed, however, that \eqref{23} involves $M_e=2m^*$, rather than the bipolaron mass.
This resolves the problem of the low temperature of condensation which arises both in the SRP theory and in the LRP theory in which expression \eqref{23} involves the bipolaron mass ~\cite{24}-\cite{27}. Another important result is that the critical temperature $T_c$ for the parameter values considerably exceeds the gap energy $\omega_0$.

From \eqref{20}, \eqref{21} follows that:
\begin{equation}\label{24}
   \frac{ N'(\tilde{\omega})}{N}=\frac{\tilde{T}^{3/2}}{C_{bp}}
	F_{3/2}\left(\frac{\tilde{\omega}}{\tilde{T}}\right)
\end{equation}

\begin{equation}\label{25}
     \frac{ N_0(\tilde{\omega})}{N}=1-\frac{ N'(\tilde{\omega})}{N}
\end{equation}
Fig.3 shows temperature dependencies of the number of supracondensate particles $N'$ and the number of particles $N_0$ occurring in the condensate for the above-indicated parameter values $\tilde{\omega}_i$.

\begin{center}

\includegraphics{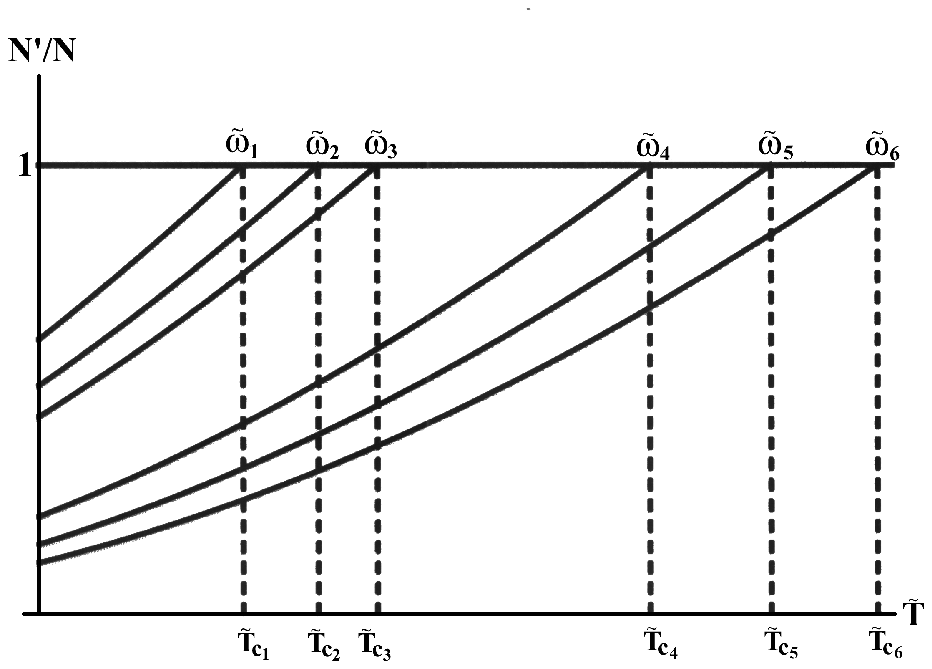}
\end{center}
\begin{center}

\includegraphics{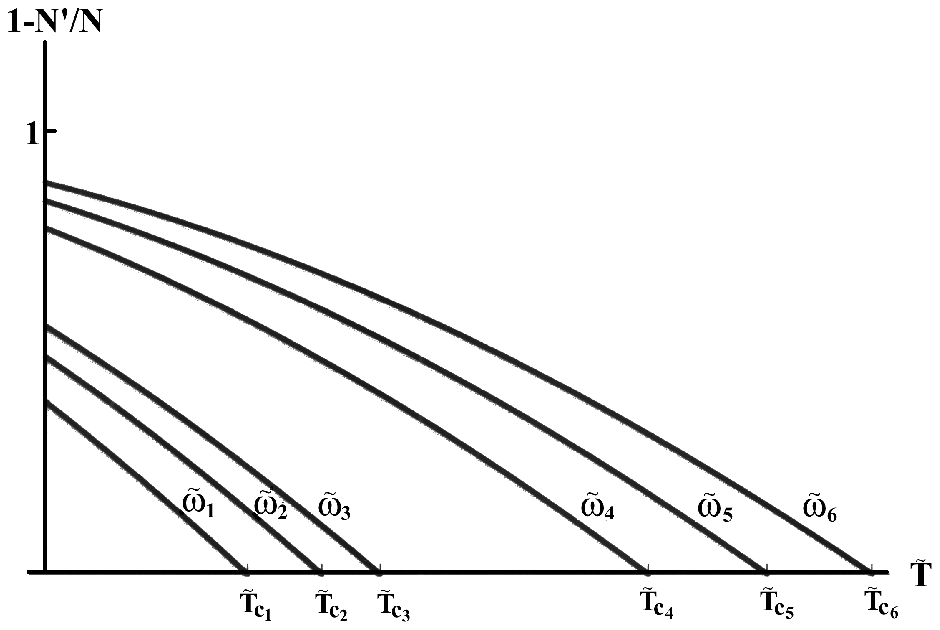}

Fig. 3. Temperature dependencies of the relative number of supracondensate particles частиц $N'/N$ and the particles occurring in the condensate $N_0/N=1-N'/N$ for the parameter values $\tilde{\omega}_i$, given in Fig.2.
\end{center}
Fig.3 suggests that, as could be expected, the number of particles in the condensate grows as the gap $\omega_i$ increases.

The energy of a TI-bipolaron gas $E$ is determined by the expression:
\begin{equation}\label{26}
    E=\sum_{n=0,1,2,...}\bar{m}_{n}E_{n}=E_{bp}N_0+\sum_{n\neq0}\bar{m}_{n}E_{n}
\end{equation}
With the use of \eqref{14},\eqref{15}, \eqref{26} the specific energy (i.e. the energy per one TI-bipolaron) $\tilde{E}(\tilde{T})=E/N\omega^*$, $\tilde{E}_{bp}=E_{bp}/\omega^*$ will be:

\begin{equation}\label{27}
    \tilde{E}(\tilde{T})=\tilde{E}_{bp}+\frac{\tilde{T}^{5/2}}{C_{bp}}
		F_{3/2}\left(\frac{\tilde{\omega}-\tilde{\mu}}{\tilde{T}}\right)
		\left[\frac{\tilde{\omega}}{\tilde{T}}+
		\frac{F_{5/2}\left(\frac{\tilde{\omega}-\tilde{\mu}}{\tilde{T}}\right)}
		{F_{3/2}\left(\frac{\tilde{\omega}-\tilde{\mu}}{\tilde{T}}\right)}\right],
\end{equation}
$$F_{5/2}(\alpha)=\frac{2}{\sqrt{\pi}}\int^{\infty}_0\frac{x^{3/2}dx}{e^{x+\alpha}-1},$$
where $\tilde{\mu}$ is determined from the equation:

\begin{equation}\label{28}
    \tilde{T}^{3/2}F_{3/2}\left(\frac{\tilde{\omega}-\tilde{\mu}}{\tilde{T}}\right)=C_{bp}
\end{equation}

$$\tilde{\mu}=\left\{\begin{array}{rl}
		0,\ \ \ \tilde{T}\leq\tilde{T}_c;\\
		\tilde{\mu}(\tilde{T}),\ \ \ \tilde{T}\geq\tilde{T}_c
		\end{array}\right. \ $$
Relation of $\tilde{\mu}$ with the chemical potential of the system $\mu$ is given by the formula $\tilde{\mu}=(\mu-E_{bp})/\omega^*$. From \eqref{27}--\eqref{28} also follow expressions for the free energy: $F=-2E/3$ and entropy $S=-\partial F/ \partial T$.

Fig.4 illustrates temperature dependencies $\Delta E=\tilde{E}-\tilde{E}_{bp}$ for the above-indicated parameter values $\omega_i$. Break points on the curves $\Delta E_i(\tilde{T})$ correspond to the values of critical temperatures $T_{c_i}$.

\begin{center}

\includegraphics{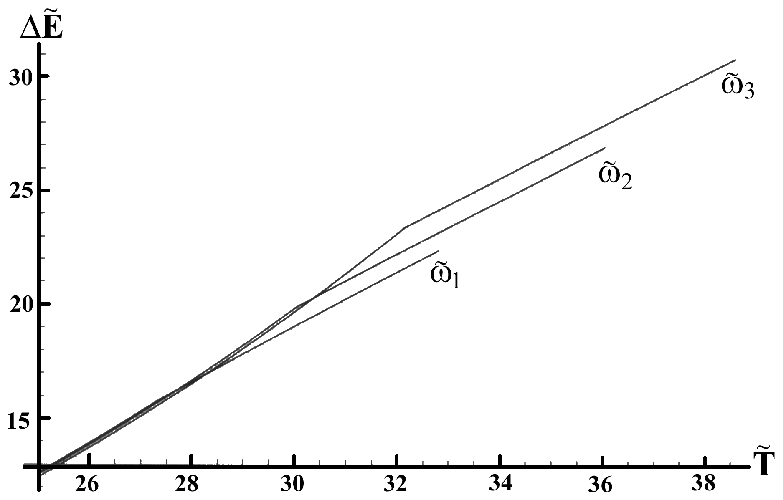}
\end{center}
\begin{center}

\includegraphics{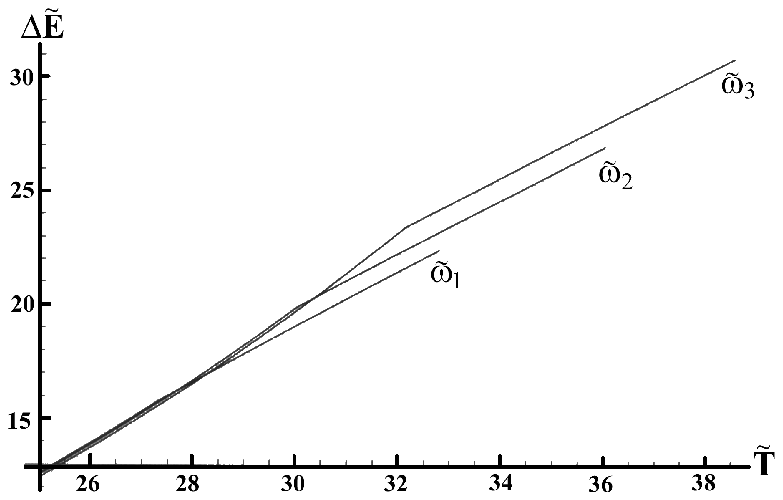}

Fig. 4. Temperature dependencies $\Delta E(\tilde{T})=\tilde{E}(\tilde{T})-\tilde{E}_{bp}$ \\
for the parameter values $\tilde{\omega}_i$ presented in Fig. 2, 3.
\end{center}
The dependencies obtained enable us to find the heat capacity of a TI-bipolaron gas: $C_v(\tilde{T})=d\tilde{E}/d\tilde{T}$.
With the use of \eqref{27} for $\tilde{T}\leq\tilde{T}_c$ we express $C_v(\tilde{T})$ as:
\begin{equation}\label{29}
    C_v(\tilde{T})=\frac{\tilde{T}^{3/2}}{2C_{bp}}
		 \left[\frac{\tilde{\omega}^2}{\tilde{T}^2}F_{1/2}\left(\frac{\tilde{\omega}}{\tilde{T}}\right)+
		 6\left(\frac{\tilde{\omega}}{\tilde{T}}\right)F_{3/2}\left(\frac{\tilde{\omega}}{\tilde{T}}\right)+
		5F_{5/2}\left(\frac{\tilde{\omega}}{\tilde{T}}\right)\right],
\end{equation}
$$F_{1/2}(\alpha)=\frac{2}{\sqrt{\pi}}\int^{\infty}_0\frac{1}{\sqrt{x}}\ \frac{dx}{e^{x+\alpha}-1}$$
Expression \eqref{29} yields a well-known exponential dependence of the heat capacity at low temperatures $C_v\backsim\text{exp}(-\omega_0/T)$, caused by the availability of the energy gap  $\omega_0$.

Fig.5 shows temperature dependencies of the heat capacity $C_v(\tilde{T})$ for the above-indicated parameter values $\tilde{\omega}_i$. Table I lists the values of the heat capacity jumps:
\begin{equation}\label{30}
    \Delta \frac{\partial C_v(\tilde{T})}{\partial\tilde{T}}=
		\left. \frac{\partial C_v(\tilde{T})}{\partial\tilde{T}}\right|_{\tilde{T}=\tilde{T}_c+0}-
		\left. \frac{\partial C_v(\tilde{T})}{\partial\tilde{T}}\right|_{\tilde{T}=\tilde{T}_c-0}
\end{equation}
at the transition points for the parameter values $\tilde{\omega}_i$.

The dependencies obtained will enable one to find the latent heat of transition $q=TS$, where $S$ is the entropy of supracondensate particles. At the point of transition this value is: $q=2T_cC_v(T_c-0)/3$, where $C_v(T)$ is determined by formula \eqref{29}. For the above-indicated parameter values $\omega_i$, it is given in Table I.

\begin{center}

\includegraphics{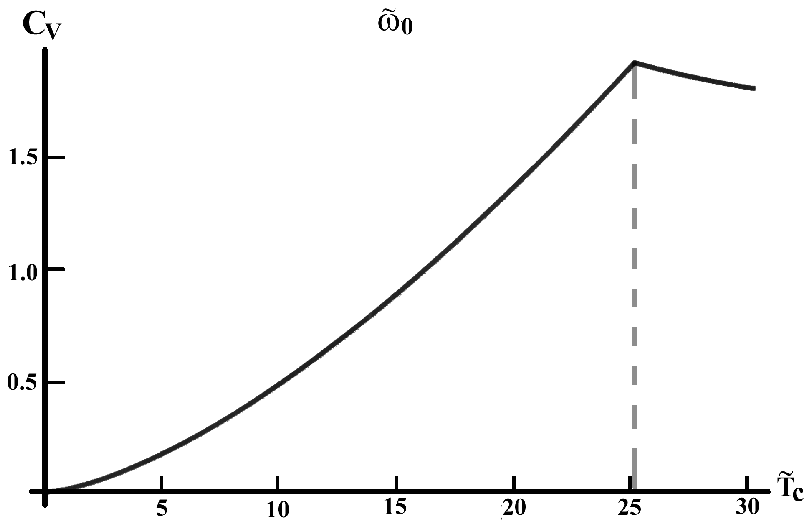}
\end{center}

{\raggedright
\includegraphics{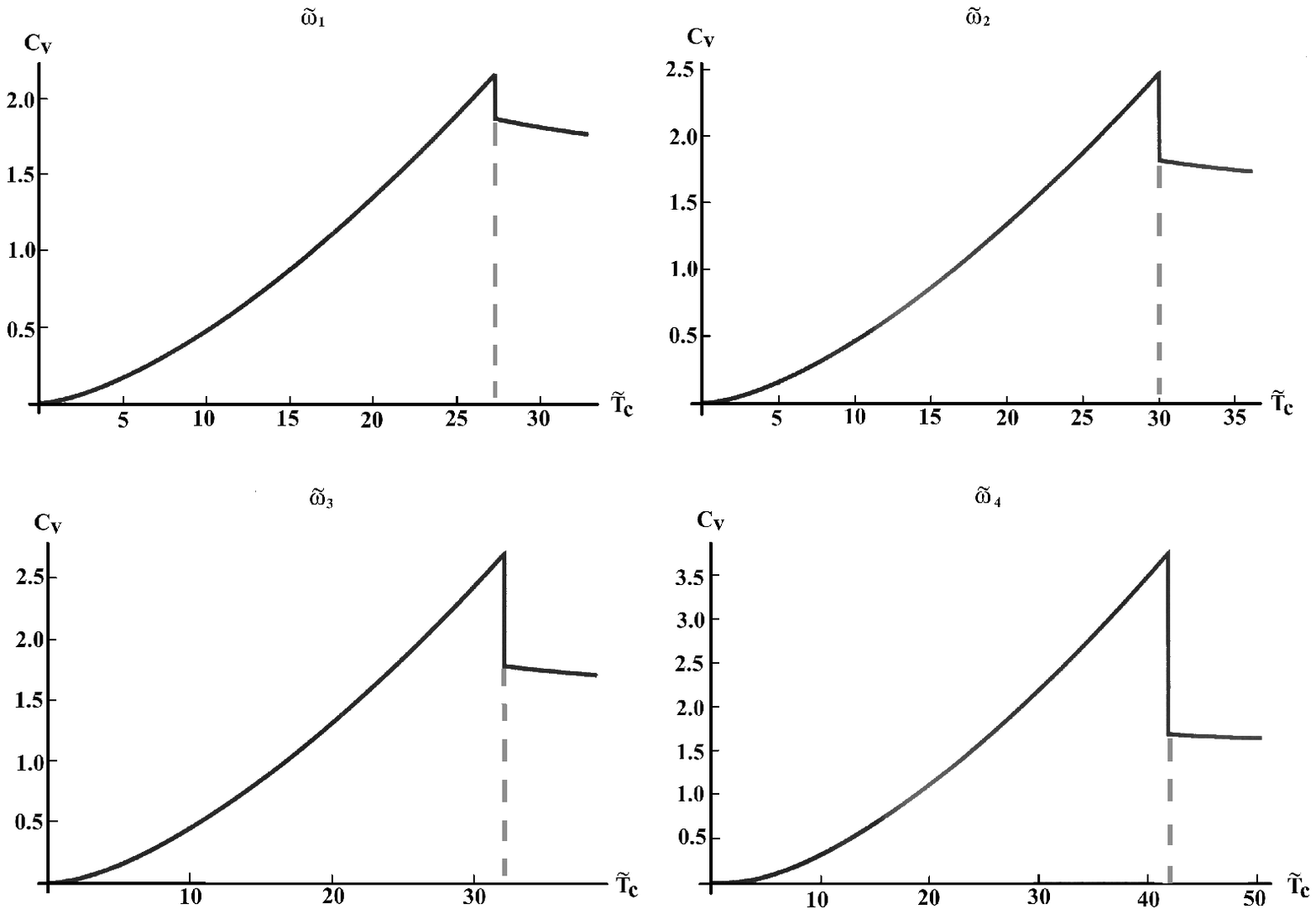}

}
\begin{center}
\includegraphics{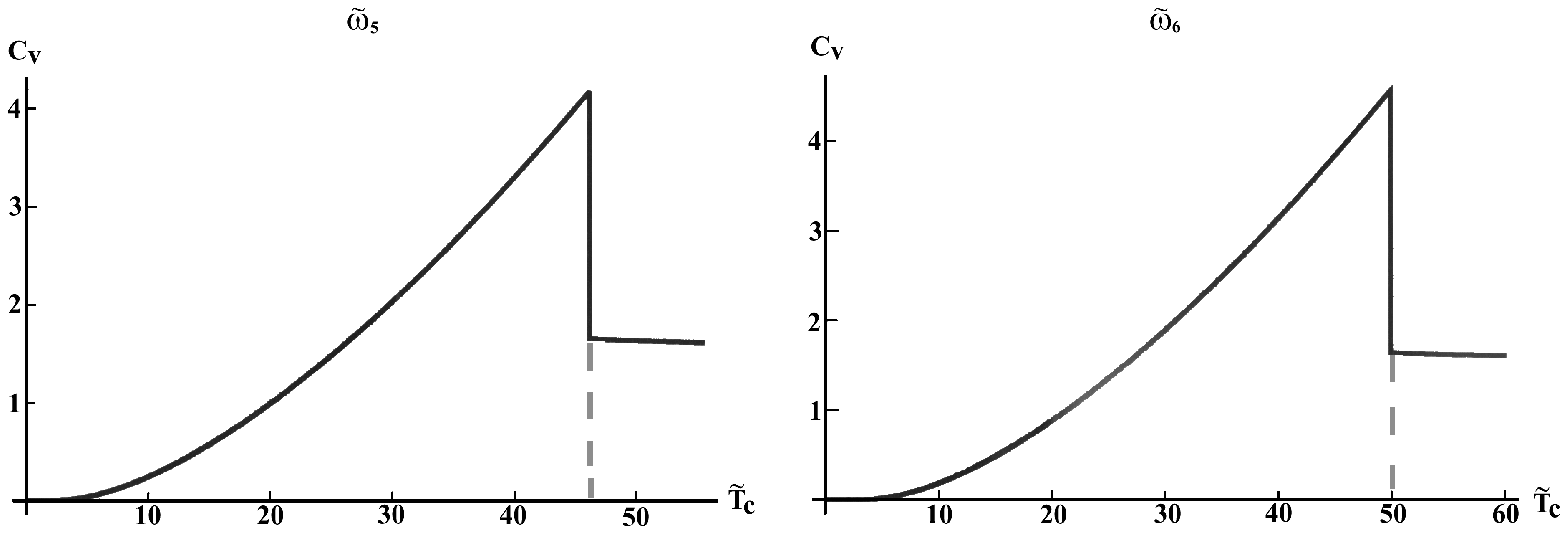}
\end{center}
\begin{center}

Fig. 5. Temperature dependencies of the heat capacity for various \\
values of the parameters $\omega_i$: $\omega_0=0$;   $\ \tilde{T}_{c0}=25,2445$; $\ C_v(\tilde{T}_{c0})=1,925$;

$\omega_1=0,2$; $\ \tilde{T}_{c1}=27,325$;  $\ C_v(\tilde{T}_{c1}-0)=2,162$;  $\ C_v(\tilde{T}_{c1}+0)=1,868$;

$\omega_2=1$;   $\ \ \tilde{T}_{c2}=30,0255$; $\ C_v(\tilde{T}_{c2}-0)=2,465$;  $\ C_v(\tilde{T}_{c2}+0)=1,812$;

$\omega_3=2$;   $\ \ \tilde{T}_{c3}=32,1397$; $\ C_v(\tilde{T}_{c3}-0)=2,699$;  $\ C_v(\tilde{T}_{c3}+0)=1,777$;

$\omega_4=10$;  $\ \tilde{T}_{c4}=41,8727$; $\ C_v(\tilde{T}_{c4}-0)=3,740$;  $\ C_v(\tilde{T}_{c4}+0)=1,6779$;

$\omega_5=15$;  $\ \tilde{T}_{c5}=46,1863$; $\ C_v(\tilde{T}_{c5}-0)=4,181$;  $\ C_v(\tilde{T}_{c5}+0)=1,651$;

$\omega_6=20$;  $\ \tilde{T}_{c6}=49,9754$; $\ C_v(\tilde{T}_{c6}-0)=4,560$;  $\ C_v(\tilde{T}_{c6}+0)=1,633$.
\end{center}

\section{Current states of a TI-bipolaron gas.}

In the foregoing we have considered equilibrium properties of a TI-bipolaron gas. The formation of Bose-condensate per se does not mean that it has superconducting properties. To demonstrate such a possibility let us consider the total momentum of a TI-bipolaron:
\begin{equation}\label{31}
    \vec{\mathcal{P}}=\hat{\vec{P}}_1+\hat{\vec{P}}_2+\sum\vec{k}a^+_ka_k,
\end{equation}
where $\hat{\vec{P}}_1$ and $\hat{\vec{P}}_2$ are the momenta of the first and second electron, respectively. It is easy to check that $\vec{\mathcal{P}}$ commutes with Hamiltonian \eqref{1} and therefore is a constant value, i.e. c-number.

For this reason, to consider nonequilibrium properties and, particularly current states we can use generalization of Heizenberg transformation \eqref{4} such that:
\begin{equation}\label{32}
    S_1(\mathcal{P})=\text{exp}\left\{i\left(\vec{\mathcal{P}}-\sum_k\vec{k}a^+_ka_k\right)\right\}\vec{R}.
\end{equation}
The general expression for the functional of the total energy of a TI-bipolaron for $\vec{\mathcal{P}}\neq 0$ is given in ~\cite{29}. TI-bipolarons occurring in condensed state have a common wave function for the whole condensate and do not thermalize  in the state when $\vec{\mathcal{P}}\neq 0$. In the supracondensate part, phonons whose wave vectors contribute into $\vec{\mathcal{P}}$ \eqref{31} are thermalized and their total momentum will be equal to zero.

Hence, all the changes arising in considering the case of $\vec{\mathcal{P}}\neq 0$ concern only the expression for the ground state energy which become dependent of $\vec{\mathcal{P}}$, while the spectrum of excited states \eqref{14}, \eqref{15} remains unchanged. It follows that in passing over the critical point, the currentless state suddenly becomes current which is in agreement with the experiment.

\section{Comparison with the experiment.}

Fig. 4 shows typical dependencies of $E(\tilde{T})$.
They suggest that at the point of transition the energy is a continuous function of $\tilde{T}$.
This means that the transition per se occurs without energy expenditure being a phase transition of the 2-kind in complete agreement with the experiment. At the same time transition of Bose particles from a condensate state to a supracondensate one occurs with consumption of energy which is determined by the value $q$ (\S4, Table 1), determining the latent heat of transition of a Bose gas which makes it a phase transition of the 1-st kind.

By way of example let us consider HTSC $YBa_2Cu_3O_7$ with the temperature of transition $90\div93$K,
volume of the unit cell $0,1734\cdot 10^{-21}$ cm$^{-3}$, concentration of holes $n\approx 10^{21}$ cm$^{-3}$.
According to estimates ~\cite{39}, Fermi energy is equal to: $\epsilon_F=0,37$ eV.
Concentration of TI-bipolarons in $YBa_2Cu_3O_7$ is found from equation \eqref{22}:

\begin{equation}\label{33}
    \frac{n_{bp}}{n}C_{bp}=f_{\tilde{\omega}}(\tilde{T}_c)
\end{equation}
with $\tilde{T}_c=1,6$. Table I lists the values of $n_{bp,i}$ for the values of parameters $\tilde{\omega}_i$ given in \S4. It follows from Table I that $n_{bp,i}<<n$.
Hence, only a small part of charge carriers is in a bipolaron state which justifies the approximation of a low-density TI-bipolaron gas used by us. The energy levels of such TI-bipolarons lie near Fermi surface and are described by the wave function:
\begin{equation}\label{34}
    \Psi(\vec{r})=e^{i\vec{k}_F\vec{r}}\varphi(\vec{r}),
\end{equation}
which leads to replacement of $\bar{T}$, involved in \eqref{6}, by:
\begin{equation}\label{35}
    \bar{T}=-2\left\langle \varphi\left|\Delta_r\right|\varphi\right\rangle+2k^2_F,
\end{equation}
that is to reckoning  of the energy from Fermi level (the last term in the right-hand side of \eqref{35} in dimensional units is equal to $2\epsilon_F$, where $\epsilon_F=\hbar^2k^2_F/2m^*$).

According to our approach, superconductivity arises when coupled states are formed. The condition for the formation of such states has the form:
\begin{equation}\label{36}
    \left|E_{bp}\right|\geq 2\left|E_{p}\right|,
\end{equation}
where $E_p$ is the energy of a TI-polaron ~\cite{34}. Condition \eqref{36} determines the value of a pseudogap:
\begin{equation}\label{37}
    \Delta _1=\left|E_{bp}-2E_p\right|
\end{equation}
For $\Delta_1>>\omega_0$ the value of a pseudogap can greatly exceed both $T_c$, and the energy of the gap (i.e. $\omega_0$).
The expression for the spectrum $E_{bp}$ and $E_p$ \eqref{14}--\eqref{15} suggests that the angular dependence $\Delta_1(\vec{k})$ is completely determined by the symmetry of the isoenergetic surface of the phonon wave vector $\vec{k}$. Earlier this conclusion was made by Bennet ~\cite{40}, who proved that the main source of anisotropy of superconducting properties is the angular dependence of the phonon spectrum, though some contribution is also made by the anisotropy of Fermi surface.

It follows from what has been said that formation of a pseudogap is a phase transition preceding the phase transition to the superconducting state. Recent experiments ~\cite{41} also testify in favor of this statement.

In paper ~\cite{32} correlation length for TI-bipolarons was calculated. According to ~\cite{32}, in HTSC materials its value can vary from several angstroms to several tens of angstroms, which is also in agreement with the experiment.

Of special interest is to determine the characteristic energy of phonons responsible for the formation of TI-bipolarons and superconducting properties of oxide ceramics. To do this let us compare the calculated values of the heat capacity jumps with experimental data.

As is known, in BCS theory a jump in the heat capacity is equal to:

$$\left.\frac{C_s-C_n}{C_n}\right|_{T_c}=1,43$$
where $C_s$ is the heat capacity in the superconducting phase, and $C_n$ is the heat capacity in the normal phase, and is independent of the parameters of the model Hamiltonian. As it follows from numerical calculations shown in fig.5 and in Table I, as distinct from the BCS theory, the value of the jump depends on the phonon frequency. Hence, the approach presented predicts the existence of an isotopy effect for the heat capacity jump.

\begin{center}

\includegraphics{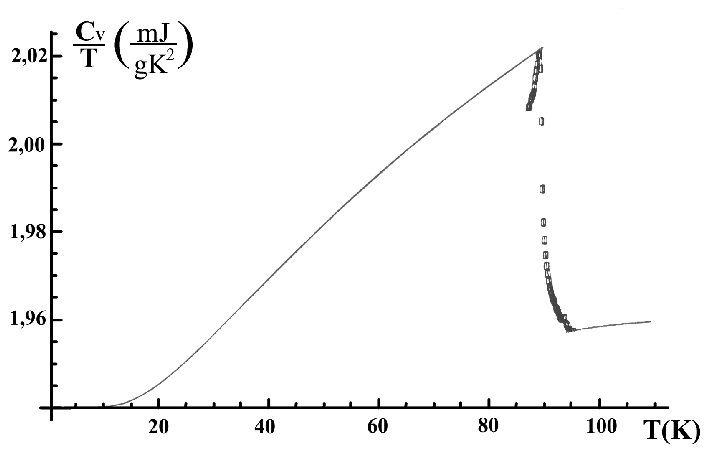}

Fig.6. Comparison of the theoretical (solid line) and experimental ~\cite{42} (broken line) dependencies in the region of the heat capacity jump.
\end{center}
As it is seen from Fig.6, the heat capacity jump calculated theoretically (\S4) coincides with the experimental value in $YBa_2Cu_3O_7$ ~\cite{42}, for $\tilde{\omega}=1,5$, i.e. for $\omega=7,5$ meV.
This corresponds to the concentration of TI-bipolarons equal to $n_{bp}=2,6\cdot 10^{18}$ cm$^{-3}$.
Hence, in contrast to the widespread notion that in oxide ceramics superconductivity is determined by high-energy phonons (with energy $~70\div80$meV ~\cite{43}) actually, the superconductivity in HTSC materials should be determined by soft phonon modes.

Notice that in calculations of the temperature of transition it was believed that the effective mass $M_e$ in equation \eqref{22} is independent of the direction of the wave vector, i.e. isotropic case was dealt with.

In the anisotropic case, choosing principal axes of vector $k$, as coordinate axes, we will get the quantity $(M_{ex}M_{ey}M_{ez})^{1/3}$ instead of the effective mass $M_e$. In complex HTSC materials the values of effective masses lying in the plane of layers $M_{ex}$, $M_{ey}$ are close in value. Assuming in this case $M_e=M_{ex}=M_{ey}=M_{\parallel}$, $M_{ez}=M_{\perp}$, we will get instead of $C_{bp}$, determined by \eqref{22}, the value $\tilde{C}_{bp}=C_{bp}/\gamma$, $\gamma^2=M_{\perp}/M_{\parallel}$ is the parameter of anisotropy. Hence anisotropy of effective masses gives for the concentration $n_{bp}$ the value $\tilde{n}_{bp}=\gamma n_{bp}$. Therefore taking account of anisotropy can an order of magnitude and greater enhance the estimate of the concentration of TI-bipolarons. If for $YBa_2Cu_3O_7$ we take the estimate $\gamma^2=30$ ~\cite{43}, then for the concentration of TI-bipolarons we will get: $\tilde{n}_{bp}=1,4\cdot 10^{19}$ cm$^{-3}$, which holds valid the general conclusion: in the case under consideration only a small number of charge carriers are in TI-bipolaron state. The situation can change if the anisotropy parameter is very large. Thus, for example, in layered HTSC Bi-Sr-Ca-Cu-O the anisotropy parameter is $\gamma>100$, accordingly, the concentration of TI-bipolarons in these compounds can have the same order of magnitude as the total concentration of charge carriers.

Another important conclusion emerging from taking account of the anisotropy of effective masses is that the temperature of the transition $T_c$ depends not on $n_{bp}$ and $M_{\parallel}$ individually, but on their relation which straightforwardly follows from \eqref{22}.

\section{Essential generalizations of the theory.}

In the foregoing we considered the case of an ideal TI-bipolaron gas. At small concentration of TI-bipolarons their Coulomb interaction will be greatly screened which justifies the use of the model of an ideal gas.

If the concentration of TI-bipolarons is large (for example $n=10^{21}$cm$^{-3}$, as was believed in \S4), then such Bose gas can no longer be considered to be ideal. Taking account of Coulomb interaction between bipolarons becomes necessary. For this purpose we can use Bogolyubov theory ~\cite{44} for weakly imperfect Bose gas, which implies that the spectrum of elementary excitations with the momentum $k$ will be determined by the expression:
\begin{equation}\label{38}
    E(k)=\sqrt{k^2u^2(k)+(k^2/2m_B)^2}
\end{equation}
where $u(k)=\sqrt{n_{B}V(k)/m_B}$, $V(k)$ is the Fourier component of the potential of pairwise interaction between charged bosons:
$V(k)=4\pi e_B^2/\epsilon_0k^2$, $e_B$ and $m_B$ are the charge and mass of a boson,
$n_B$ is the concentration of bosons.

Whence it follows that:
\begin{equation}\label{39}
    E(k)=\sqrt{(\hbar\omega_p)^2+(k^2/2m_B)^2},\ \ \
		\omega_p=\sqrt{4\pi e^2n_{B}/\epsilon_0m_B}
\end{equation}
$\omega_p$ is the plasma frequency. According to (39), the spectrum of excitations of quasiparticles of an imperfect gas is characterized by a finite energy gap which in the long-wavelength limit is equal to $\omega_p$. The same result can be arrived at if the Coulomb interaction between the electrons is considered as a result of electron-plasmon interaction. According to ~\cite{45}, Hamiltonian of electron-plasmon interaction coincides in structure with Froehlich Hamiltonian which involves plasmon frequency instead of phonon frequency. This straightforwardly leads to the energy gap equal to the plasmon frequency $\omega_p$, if in (39) we put: $e_B=2e$, $m_B=2m^*$, $n_B=n/2$, where $n$ is the electron concentration. Hence, for $\omega_p<\omega_0$ the energy gap will be determined by $\omega_p$, while for $\omega_p>\omega_0$ it will still be determined by $\omega_0$.

Actually, in real HTSC, there are not only phonon and plasmon branches, but also some other elementary excitations which can take part in electron pairing. An example is spin fluctuations. Generalization of the theory to the case of interaction with various brunches of excitations which will contribute into the ground state energy, the value of the gap and the dispersion law of a TI-bipolaron is a topical problem. Obviously, taking account of this interaction will lead to an increase in the coupling energy of both TI-bipolarons and TI-polarons. Therefore, a priori, without any particular calculations one cannot say anything of how the condition of stability of TI-bipolaron states determined by \eqref{36} will change.

Another important problem is generalization of the theory to the case of intermediate coupling of electron-phonon interaction. Formally, the expression for the functional of the ground state energy of a TI-bipolaron \eqref{12} is valid for any value of the electron-phonon coupling constant. For this reason such a calculation will not change the spectrum of a TI-bipolaron, however it will alter the criteria of fulfillment of the conditions of a TI-bipolaron stability \eqref{36}.

\section{Conclusive remarks.}

In this paper we have presented conclusions emerging from consistent translation-invariant consideration of EPI. It implies that, pairing of electrons, for any coupling constant, leads to a concept of TI-polarons and TI-bipolarons. Being bosons, TI-bipolarons can experience Bose condensation leading to superconductivity. Let us list the main results following from this approach. First and foremost the theory resolves the problem of the great value of the bipolaron effective mass (\S4). As a consequence, formal limitations on the value of the critical temperature of the transition are eliminated too. The theory quantitatively explains such thermodynamic properties of HTSC-conductors as availability (\S4) and value (\S6) of the jump in the heat capacity lacking in the theory of Bose condensation of an ideal gas.
The theory also gives an insight into the occurrence of a great ratio between the width of the pseudogap and $T_c$ (\S6). It accounts for the small value of the correlation length~\cite{32} and explains the availability of a gap (\S3) and a pseudogap (\S6) in HTSC materials. The angular dependence of the gap and pseudogap gets a natural explanation (\S6).

Accordingly, isotopic effect automatically follows from expression \eqref{22}, where the phonon frequency $\omega_0$ acts as a gap. The conclusion of the dependence of the temperature of the transition $T_c$ on the relation $n_{bp}/M_{\parallel}$ (\S6) correlates with Uemura law universal for all HTSC materials which implies that the temperature of the transition relates to the concentration of charge carriers divided by its effective mass ~\cite{46}.

Application of the theory to 1D and 2D systems leads to qualitatively new results since the occurrence of a gap in the TI-bipolaron spectrum automatically removes divergences at small momenta, inherent in the theory of ideal Bose gas. These problems will be considered by the author in a special paper. The case of 1D-polaron was discussed in~\cite{49}.

The work was done with the support from the RFBR, project N 13-07-00256.

\section{Appendix. Remarks on the notation.}

Function $F_{3/2}(\alpha)$ is called a polylogarithm $=Li_{3/2}(e^{-\alpha})$,
in mathematics this is the function $PolyLog$, therefore the function $f_{\tilde{\omega}}$ in \eqref{22} will be: $f_{\tilde{\omega}}=\tilde{T}^{3/2}PolyLog\left[3/2, e^{-\tilde{\omega}/\tilde{T}}\right]$.

In the general case, function $PolyLog$ of the order of $s$ is determined as:
$$PolyLog\left[s,e^{-\alpha}\right]=\frac{1}{\Gamma(s)}\int^{\infty}_0\frac{t^{s-1}}{e^{t+\alpha}-1}dt$$
where $\Gamma(s)$ is a gamma function: $\Gamma(1/2)=\sqrt{\pi}$, $\Gamma(3/2)=\sqrt{\pi}/2$, $\Gamma(5/2)=3\sqrt{\pi}/4$.

Accordingly, the functions $F_{1/2}$, $F_{3/2}$, $F_{5/2}$ occurring in the text will be:

$F_{1/2}=2PolyLog\left[1/2, e^{-\tilde{\omega}/\tilde{T}}\right]$;

$F_{3/2}=PolyLog\left[3/2, e^{-\tilde{\omega}/\tilde{T}}\right]$;

$F_{5/2}={3/2}PolyLog\left[5/2, e^{-\tilde{\omega}/\tilde{T}}\right]$.

\newpage

\begin {center} {Table I. Calculated characteristics of Bose-gas of TI-bipolarons with concentration $n=10^{21}$ cm$^{-3}$}.
\end {center}

\begin{table}[h]
	\centering
		\begin{tabular}{|c|c|c|c|c|c|c|c|} \hline
		i & 0 & 1 & 2 & 3 & 4 & 5 & 6  \\ \hline
		$\tilde{\omega}_i$ & 0 & 0,2 & 1 & 2 & 10 & 15 & 20 \\ \hline
		$T_{ci}/\omega_{oi}$ & $\infty$ & 136,6 & 30,03 & 16,07 & 4,187 & 3,079 & 2,499 \\ \hline
		$q_i/T_{ci}$& 1,28 & 1,44 & 1,64 & 1,80 & 2,49 & 2,79 & 3,04 \\ \hline
		$-\Delta(\partial C_{v,i}/\partial\tilde{T})$ & 0,114 & 0,119 & 0,124 & 0,128 & 0,143 & 0,148 & 0,152 \\ \hline
		$C_{v,i}(T_c-0)$ & 1,925 & 2,162 & 2,465 & 2,699 & 3,740 & 4,181 & 4,560 \\ \hline
		$(C_s-C_n)/C_n$ & 0 & 0,1574 & 0,3604 & 0,519 & 1,229 & 1,532 & 1,792 \\ \hline
		$n_{bp_i}\cdot$cm$^{3}$ & 16$\cdot 10^{19}$ & 9,4$\cdot 10^{18}$ & 4,2$\cdot 10^{18}$ & 2,0$\cdot 10^{18}$ &
		1,2$\cdot 10^{17}$ & 5,2$\cdot 10^{14}$ & 2,3$\cdot 10^{13}$ \\ \hline
		\end{tabular}
\end{table}

$\tilde{\omega}_i=\omega_i/\omega^*$, $\omega^*=5$meV, $\omega_i$ is the energy of an optical phonon;
$T_{c_i}$ is the critical temperature of the transition,
$q_i$ is the latent heat of the transition from condensate to supracondensate state;
$-\Delta(\partial C_{v,i}/\partial\tilde{T})=\left.\partial C_{v,i}/\partial\tilde{T}\right|_{\tilde{T}=\tilde{T}_{c_i}+0}
-\left.\partial C_{v,i}/\partial\tilde{T}\right|_{\tilde{T}=\tilde{T}_{c_i}-0}$ is the jump in heat capacity during SC transition,
$\tilde{T}=T/\omega^*$ ; $C_{v,i}(T_c-0)$ is the heat capacity in the SC phase at the critical point;
$C_s=C_v(T_c-0)$, $C_n=C_v(T_c+0)$;
The calculations are carries out for the concentration of TI-bipolarons $n=10^{21}$ cm$^{-3}$ and the effective mass of a band electron $m^*=m_0$.

The table also lists the values of concentrations of TI-bipolarons $n_{bpi}$ for HTSC $YBa_2Cu_3O_7$, based on the experimental value of the transition temperature $T_c=93K$ (\S6).

\end{document}